\documentclass[twocolumn,superscriptaddress,nobalancelastpage,10pt,pra,a4paper]{revtex4}
\usepackage{amsmath}
\usepackage{amsthm}
\usepackage{amssymb}
\usepackage{amsfonts}
\usepackage{graphicx}
\usepackage{wasysym}
\usepackage{mathrsfs}
\usepackage{yfonts}
\usepackage{bbold}
\usepackage{verbatim}
\usepackage{subfigure}
\usepackage{color}

\newcommand{\ket}[1]{\left| #1 \right\rangle}
\newcommand{\bra}[1]{\left\langle #1 \right|}

\newcommand{\opa}{\widehat{a}}

\newcommand{\opEps}{\widehat{E}^{(+)}}
\newcommand{\opEns}{\widehat{E}^{(-)}}

\renewcommand{\r}{\mathbf{r}}
\renewcommand{\k}{\mathbf{k}}

\makeatletter
\newcommand*{\rom}[1]{\expandafter\@slowromancap\romannumeral #1@}
\makeatother

\DeclareMathAlphabet{\mathpzc}{OT1}{pzc}{m}{it}

\begin{document}





\title{Twin Photon Correlations in Single-Photon Interference}

\author{Mayukh Lahiri}
\email{mayukh.lahiri@univie.ac.at} \affiliation{Vienna Center for
Quantum Science and Technology (VCQ), Faculty of Physics, University
of Vienna, Boltzmanngasse 5, Vienna A-1090, Austria.}

\author{Armin Hochrainer} \affiliation{Vienna Center for
Quantum Science and Technology (VCQ), Faculty of Physics, University
of Vienna, Boltzmanngasse 5, Vienna A-1090, Austria.}

\author{Radek Lapkiewicz} \affiliation{Faculty of Physics,
University of Warsaw, Pasteura 5, 02-093 Warsaw, Poland.}

\author{Gabriela Barreto Lemos} \affiliation{Vienna Center for
Quantum Science and Technology (VCQ), Faculty of Physics, University
of Vienna, Boltzmanngasse 5, Vienna A-1090,
Austria.}\affiliation{Institute for Quantum Optics and Quantum
Information, Austrian Academy of Sciences, Boltzmanngasse 3, Vienna
A-1090, Austria.}

\author{Anton Zeilinger}\affiliation{Vienna Center for
Quantum Science and Technology (VCQ), Faculty of Physics, University
of Vienna, Boltzmanngasse 5, Vienna A-1090,
Austria.}\affiliation{Institute for Quantum Optics and Quantum
Information, Austrian Academy of Sciences, Boltzmanngasse 3, Vienna
A-1090, Austria.}

\begin{abstract}
We show that it is possible to generate a novel single-photon fringe
pattern by using two spatially separated identical bi-photon
sources. The fringes are similar to the ones observed in a Michelson
interferometer and possess certain remarkable properties with
potential applications. A striking feature of the fringes is that
although the pattern is obtained by detecting only one photon of
each photon pair, the fringes shift due to a change in the optical
path traversed by the undetected photon. The fringe shift is
characterized by a combination of wavelengths of both photons, which
implies that the wavelength of a photon can be measured without
detecting it. Furthermore, the visibility of the fringes diminishes
as the correlation between the transverse momenta of twin photons
decreases: visibility is unity for maximum momentum correlation and
zero for no momentum correlation. We also show that the momentum
correlation between the two photons of a pair can be determined from
the single-photon interference pattern. We thus for the first time
propose a method of measuring a two-photon correlation without
coincidence or heralded detection.
\end{abstract}

\maketitle

\section{Introduction}\label{sec:intro}
The discovery of the Michelson interferometer
\cite{Mich-interf-orig} is one of the most important events in the
history of physics: apart from its relevance to the special theory
of relativity \cite{MM-rel-exp,E-sp-rel}, it has also been applied
in matter-wave interferometry \cite{Schmiedmayer-RMP}, and, most
recently, in the detection of gravitational waves \cite{Ligo-PRL}.
Here we establish that a novel type of single-photon fringe pattern
can be created, which looks similar to the one observed in a
Michelson interferometer, but possesses some remarkable properties.
We produce the fringes using the method of ``induced coherence
without induced emission'' \cite{ZWM-ind-coh-PRL,WZM-ind-coh-PRA}.
The method is based on the following quantum mechanical principle:
quantum interference occurs if and only if the information regarding
the path traversed by a quantum entity is unavailable
\cite{Feynman-lec-3}. This method has already been applied to the
areas of imaging \cite{LBCRLZ-mandel-im,LLLZ-th-mandel-img},
spectroscopy \cite{Kulik-spec-2004,Kulik-spec-2016}, optical
polarization \cite{LHLBZ-quant-pol}, tests of the complementarity
principle
\cite{HRWZ-frust-phot,HKWZ-comp-era,HMM-comp-two-ph-PRL,HMM-comp-two-ph-PRA},
and microwave superconducting cavities \cite{LPHH-mandel-MSC}.
\par
In a Michelson interferometer, the two interfering beams are
produced from an original beam by the method of division of
amplitude and the fringe shift associated with a change in optical
path is characterized by the wavelength of the interfering beams
(\cite{BW}, Sec. 7.5.4). By contrast, the interfering beams in our
case are produced by two spatially separated identical sources each
of which generates photon pairs; the fringe shift associated with a
change in the optical path is characterized by the wavelengths of
both photons that constitutes a pair. This fact can be used to
determine the wavelength of a photon without detecting that photon.
Furthermore, the visibility of these fringes depends on the
correlation between the transverse momenta of the two photons; in
certain cases, this fact allows us to quantitatively determine the
momentum correlation between the two photons belonging to a photon
pair by detecting only one of the photons.
\par
In Sec. \ref{sec:back-theory}, we give a summary of the notations to
be used in this paper. In Sec. \ref{sec:theory-interf}, we present
the main theoretical analysis and discuss the properties of the
fringes in detail. Then in Sec. \ref{sec:mom-cor-from-vis}, we show
that under certain reasonable assumptions it is possible to obtain a
measure of the momentum correlation between twin photons from the
visibility of the fringe pattern. After that, in Sec.
\ref{sec:exp-confirm}, we briefly compare the theoretical
predictions with experimental observations. Finally, in Sec.
\ref{sec:conc}, we summarize our results and discuss their
implications.

\section{Notations}\label{sec:back-theory}
We assume that the two photons, $a$ and $b$, constituting a pair
have, in general, different values of mean frequency (energy). If
the associated optical fields are distributed over several
plane-wave modes (spatial modes), the quantum state of the photon
pair can be represented in the form \cite{Note-cont-var}
\begin{align}\label{state-two-ph-mm}
\ket{\psi}=\sum_{\k_{a},\k_{b}}C_{\k_{a},\k_{b}} \ket{\k_{a}}_{a}
\ket{\k_{b}}_{b},
\end{align}
where $\ket{\k_{a}}_{a}$ denotes a single $a$-photon occupation in
the mode labeled by the wave vector $\k_{a}$ and the complex
amplitudes $C_{\k_{a},\k_{b}}$ assure that the state $\ket{\psi}$ is
normalized.
\par
The joint probability (density) of photon $a$ having momentum
$\pmb{\mathpzc{p}}_a=\hbar\k_{a}$ and photon $b$ having momentum
$\pmb{\mathpzc{p}}_b=\hbar\k_{b}$ is equal to
\begin{align}\label{joint-prob}
P(\k_{a},\k_{b})=|C_{\k_{a},\k_{b}}|^2.
\end{align}
The conditional probability of photon $a$ to have momentum
$\hbar\k_{a}$ given photon $b$ carries momentum $\hbar\k_{b}$ is
given by
\begin{align}\label{cond-prob}
\mathcal{P}(\k_{a}|\k_{b})\equiv\frac{P(\k_{a},\k_{b})}{P(\k_b)}
=\frac{|C_{\k_{a},\k_{b}}|^2}{\sum_{\k_a}|C_{\k_{a},\k_{b}}|^2},
\end{align}
where
$P(\k_b)=\sum_{\k_a}P(\k_{a},\k_{b})=\sum_{\k_a}|C_{\k_{a},\k_{b}}|^2$
is the probability of photon $b$ having momentum $\hbar\k_{b}$.
Similarly, $\mathcal{P}(\k_{b}|\k_{a})
=|C_{\k_{a},\k_{b}}|^2/P(\k_a)$ and
$P(\k_a)=\sum_{\k_b}|C_{\k_{a},\k_{b}}|^2$.
\par
The correlation between $\pmb{\mathpzc{p}}_a$ and
$\pmb{\mathpzc{p}}_b$ is governed by $P(\k_{a},\k_{b})$. For
example, when $\pmb{\mathpzc{p}}_a$ and $\pmb{\mathpzc{p}}_b$ are
fully uncorrelated (statistically independent), the conditional
probability $\mathcal{P}(\k_{a}|\k_{b})=P(\k_{a})$; in this case the
joint probability takes the form $P(\k_{a},\k_{b})=P(\k_a)P(\k_b)$.
It is thus clear from Eq. (\ref{state-two-ph-mm}) that if the
quantum state can be expressed in the product form
$\ket{\psi}=\ket{\psi_a}\otimes \ket{\psi_b}$, the momenta of the
photons are uncorrelated.
\par
Throughout this paper we assume that the photons propagate as
paraxial beams and are incident normally on a detector. Therefore,
the correlation between momenta is to be understood as correlation
between \emph{transverse} momenta of the photons.

\section{Single-photon Interference Using Two Identical Biphoton
Sources}\label{sec:theory-interf}

\begin{figure}[htbp]  \centering
\includegraphics[width=0.25\textwidth]{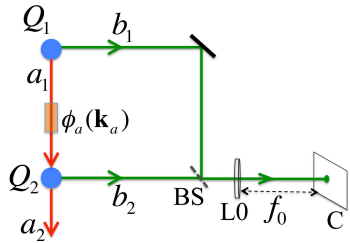}
\qquad \caption{Schematic of the proposed experiment. Biphoton
sources $Q_j$ emit photons $a$ and $b$ into beams $a_j$ and $b_j$.
Beam $a_1$ is sent through $Q_2$ and aligned with $a_2$. The
$b-$photon beams generated by the sources are superposed by a beam
splitter, BS, and one of the outputs of BS is focused by a positive
lens, L0, of focal length $f_0$ on a camera, C. A plane-wave
($\k_a$) mode of photon $a$ gains phase $\phi_a(\k_a)$ due to
propagation from $Q_1$ to $Q_2$.} \label{fig:a-ph-align-ed}
\end{figure}

Let $Q_1$ and $Q_2$ be \emph{identical} sources (Fig.
\ref{fig:a-ph-align-ed}), each of which generates biphoton states
given by Eq. (\ref{state-two-ph-mm}). Source $Q_j$ ($j=1,2$) emits
photons $a$ and $b$ into the beams $a_j$ and $b_j$, respectively.
The beams, $b_1$ and $b_2$, are superposed by a beam splitter, BS.
They interfere if and only if one cannot identify the path ($b_1$ or
$b_2$) traversed by photon $b$ that emerges from an output of BS. No
path information is available if one takes the two following
measures \cite{WZM-ind-coh-PRA,Note-Ou}: 1) choosing the optical
path lengths appropriately; 2) sending beam $a_1$ through $Q_2$ and
aligning it with beam $a_2$ such that the spatial modes present in
$a_2$ are identical with those present in $a_1$. We assume that the
simultaneous presence of photons generated by both sources is highly
improbable; this also implies that almost no stimulated emission
occurs at $Q_2$. Under these circumstances the quantum state of
light in the system is given by (see Appendix 1; cf.
\cite{ZWM-ind-coh-PRL,WZM-ind-coh-PRA,LLLZ-th-mandel-img})
\begin{align}\label{state-sup-prod-mm}
& \ket{\Psi}= ~\alpha_1 \sum_{\k_{a_1},\k_{b_1}}
C_{\k_{a_1},\k_{b_1}}\ket{\k_{a_1}}_{a_1} \ket{\k_{b_1}}_{b_1}
\nonumber \\& + \alpha_2 \sum_{\k_{a_2},\k_{b_2}}
\exp[-i\phi_a(\k_a)] C_{\k_{a_2},\k_{b_2}} \ket{\k_{a_2}}_{a_1}
\ket{\k_{b_2}}_{b_2},
\end{align}
where $\phi_a(\k_a)$ is the phase acquired by the plane-wave mode
$\k_a$ due to propagation from $Q_1$ to $Q_2$, $\alpha_1$ and
$\alpha_2$ are complex numbers obeying
$|\alpha_1|^2+|\alpha_2|^2=1$; $|\alpha_j|$ characterizes the rate
of emission from $Q_j$.
\par
\begin{figure}[htbp]  \centering
\includegraphics[width=0.25\textwidth]{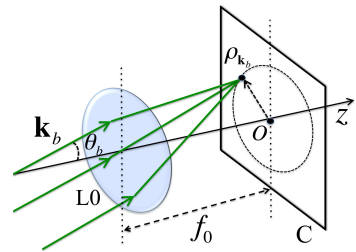}
\qquad \caption{Detection system geometry. L0 is a positive lens
with focal length $f_0$. The origin ($O$) is chosen at the point
where the optical axis (beam axis) $z$ meets the detection plane
(camera). The wave vector $\k_b$ makes an angle $\theta_b$ with the
optical axis. Plane waves making an angle $\theta_b$ with the $z$
axis are focused along a circle of radius
$|\pmb{\rho}_{\k_b}|\approx f_0\theta_b$ centered at $O$;
$\pmb{\rho}_{\k_b}$ is a two-dimensional position vector.}
\label{fig:det-sys}
\end{figure}

One of the outputs of BS is focused by a thin positive lens, L0, on
a camera, C. Within the diffraction limit, the positive lens maps a
plane wave with wave vector $\k_b$ on a point, $\pmb{\rho}_{\k_b}$,
on the camera (Fig. \ref{fig:det-sys}). The positive frequency part
of the quantized field at $\pmb{\rho}_{\k_b}$ can, therefore, be
expressed as (see also, \cite{LLLZ-th-mandel-img})
\begin{align}\label{E-quant-at-cam-fringe}
\opEps_{b}(\pmb{\rho}_{\k_b})=\opa_{b_1}(\k_{b}) +i
\exp[i\phi_b(\k_b)]\opa_{b_2}(\k_{b}),
\end{align}
where $\opa_{b_j}(\k_{b})$ is the photon-annihilation operator such
that $\opa^{\dag}_{b_j}(\k_{b})\opa_{b_j}(\k_{b})\ket{\k_{b}}_{b_l}
=\delta_{jl}\ket{\k_{b}}_{b_l}$, and $\phi_b=\phi_{b_2}-\phi_{b_1}$
is the phase difference resulting from different propagation lengths
of the beams $b_1$ and $b_2$.
Apart from a proportionality constant, the photon counting rate
\cite{G-1} at a point in the camera is given by
\begin{equation}\label{R-at-cam-def}
\mathcal{R}(\pmb{\rho}_{\k_b})\equiv
\bra{\Psi}\opEns_{b}(\pmb{\rho}_{\k_b})\opEps_{b}(\pmb{\rho}_{\k_b})
\ket{\Psi},
\end{equation}
where $\opEns_{b}(\pmb{\rho}_{\k_b})=
\{\opEps_{b}(\pmb{\rho}_{\k_b})\}^{\dag}$. It follows from Eqs.
(\ref{state-sup-prod-mm}), (\ref{E-quant-at-cam-fringe}), and
(\ref{R-at-cam-def}) that
\begin{align}\label{R-at-cam-fringe}
&\mathcal{R}(\pmb{\rho}_{\k_b})=\sum_{\k_a}|C_{\k_{a},\k_{b}}|^2
\{|\alpha_1|^2+|\alpha_2|^2 \nonumber
\\& \qquad+2|\alpha_1||\alpha_2|\cos [\phi_b(\k_b) -\phi_a(\k_a)
+\phi_2-\phi_1 ] \},
\end{align}
where $\phi_1=\text{arg}\{\alpha_1\}$,
$\phi_2=\text{arg}\{\alpha_2\}$, arg being the argument of a complex
number. Clearly, phase changes introduced by both $a$- and
$b$-photons modulate the photon counting rate.
\par
We are interested in the case where the photon counting rate at the
camera is modulated only by the phase term, $\phi_a(\k_a)$, i.e., by
the phase introduced by photon $a$; we assume that $\phi_a(\k_a)$ is
\emph{not} a slowly varying function of $\k_a$. We set the
difference between the optical path traversed by $b_1$ and $b_2$
small enough such that $\phi_b(\k_b)$ becomes a slowly varying
function of $\k_b$ and can be treated as a constant. If we also
assume that the sources emit at the same rate (i.e.,
$|\alpha_1|=|\alpha_2|$), Eq. (\ref{R-at-cam-fringe}) reduces to the
form
\begin{align}\label{R-at-cam-a-fringe-sp}
\mathcal{R}(\pmb{\rho}_{\k_b}) \propto
\sum_{\k_a}|C_{\k_{a},\k_{b}}|^2 \{1+ \cos[\phi_a(\k_a)-\phi_0] \},
\end{align}
where all other phase terms are included in $\phi_0$. Note that the
cosine term cannot be pulled out of the summation, which suggests
that several spatial modes ($\k_a$) of an $a$-photon can contribute
to the photon counting rate at a single point ($\pmb{\rho}_{\k_b}$)
on the camera. Furthermore, these contributions are weighted with
the joint probability
$\mathcal{P}(\k_{a},\k_{b})=|C_{\k_{a},\k_{b}}|^2$. The correlation
between the transverse momenta of photons $a$ and $b$ thus affects
the properties of the resulting fringe pattern observed on the
camera.
\par
In particular, we are interested in the visibility of the fringe
pattern. The visibility at a point $(\pmb{\rho}_{\k_b})$ on the
fringe pattern is defined by the usual formula \cite{BW}
\begin{align}\label{vis-def}
\mathcal{V}(\pmb{\rho}_{\k_b})
=\frac{\mathcal{R}_{\text{max}}(\pmb{\rho}_{\k_b})
-\mathcal{R}_{\text{min}}(\pmb{\rho}_{\k_b})}
{\mathcal{R}_{\text{max}}(\pmb{\rho}_{\k_b})
+\mathcal{R}_{\text{min}}(\pmb{\rho}_{\k_b})},
\end{align}
where $\mathcal{R}_{\text{max}}(\pmb{\rho}_{\k_b})$ and
$\mathcal{R}_{\text{min}}(\pmb{\rho}_{\k_b})$ are maximum and
minimum values of the photon counting rate, respectively, at the
point $\pmb{\rho}_{\k_b}$; the maximum and minimum values are
obtained by varying the phase term $\phi_0$.
\par
In the subsections below, we discuss the relationship between the
fringe visibility and momentum correlation between photons $a$ and
$b$. We consider three cases where the momenta are maximally,
minimally, and partially correlated.

\subsection{Maximal Momentum Correlation} \label{subsec:max-corr}
Suppose that photons $a$ and $b$ have mean frequencies
$\bar{\omega}_a$ and $\bar{\omega}_b$, respectively, such that the
moduli of the associated wave vectors in the vacuum are given by
$|\k_a|=\bar{\omega}_a/c$ and $|\k_b|=\bar{\omega}_b/c$, $c$ being
the speed of light in vacuum. We further assume that the beam axis
is identical with the optical axis, i.e., the symmetry axis of the
optical system.
\par
We first consider the situation in which the momenta of photons $a$
and $b$ are maximally correlated, i.e., if photon $b$ is detected in
mode $\k_b$, photon $a$ must be detected in mode
$\k_a=\mathbf{f}(\k_b)$, where the vector $\mathbf{f}(\k_b)$ is
uniquely defined for any $\k_b$. One thus has
$\mathcal{P}(\k_{a}|\k_{b})=\delta^{(3)}_{\k_a,\mathbf{f}(\k_b)}$,
where $\delta^{(3)}_{\k,\k''}=1$ for $\k=\k''$, and
$\delta^{(3)}_{\k,\k''}=0$ for $\k\neq \k''$; i.e.,
\begin{align}\label{cond-del-corr}
|C_{\k_{a},\k_{b}}|^2=P(\k_b)\delta^{(3)}_{\k_a,\mathbf{f}(\k_b)}.
\end{align}
It now follows from Eqs. (\ref{R-at-cam-a-fringe-sp}) and
(\ref{cond-del-corr}) that
\begin{align}\label{R-at-cam-a-fringe-sp-max-ent}
\mathcal{R}(\pmb{\rho}_{\k_b})\propto P(\k_b) \left\{1+
\cos\big(\phi_a[\k_a=\mathbf{f}(\k_b)]-\phi_0\big) \right\}.
\end{align}
\par
Let $d_a$ be the effective propagation distance between $Q_1$ and
$Q_2$ along the axis of the beam $a_1$. The length of the optical
path traveled along $\k_a$ that forms an angle $\theta_a$ with the
beam axis, is given by $n(\bar{\omega}_a)d_a/\cos\theta_a$; one
therefore has
\begin{align}\label{ph-ch-a-fringes}
\phi_a(\k_a)=\frac{\bar{\omega}_a}{c}\frac{n(\bar{\omega}_a)
d_a}{\cos\theta_a}\approx \frac{\bar{\omega}_a}{c}n(\bar{\omega}_a)
d_a(1+\frac{\theta_a^2}{2}),
\end{align}
where $n(\bar{\omega}_a)$ is the refractive index of the medium
between $Q_1$ and $Q_2$. By choosing an appropriate value of
$\phi_0$, it is possible to set $\bar{\omega}_a n(\bar{\omega}_a)
d_a/c-\phi_0$ equal to a multiple of $2\pi$. Equation
(\ref{R-at-cam-a-fringe-sp-max-ent}) now becomes
\begin{align}\label{R-at-cam-a-fringe-sp-max-ent-red}
\mathcal{R}(\pmb{\rho}_{\k_b})\propto P(\k_b) \{1
+\cos[\bar{\omega}_a n(\bar{\omega}_a) d_a\theta_a^2/(2c)]\}.
\end{align}
It is clear that when $d_a$ is large enough, interference fringes of
unit visibility appear on the camera \cite{Note-lens-sys}. The shape
of these fringes depends on the relationship between $\theta_a$ and
$\pmb{\rho}_{\k_b}$, i.e., on the form of $\mathbf{f}(\k_b)$.
\par
\begin{figure}[htbp]  \centering
\includegraphics[width=0.18\textwidth]{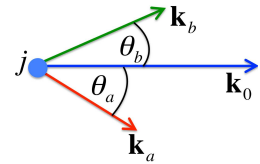}
\qquad \caption{Illustrating the case $\k_a=\mathbf{f}(\k_b)\approx
\k_0-\k_b$. The constant vector $\k_0$ is along the common axis of
$a$- and $b$-photon beams. The angle between $\k_a$ and $\k_0$ is
$\theta_a$; the angle between $\k_b$ and $\k_0$ is $\theta_b$. Both
$\theta_a$ and $\theta_b$ are small angles.}
\label{fig:th-a-b-rel-ed}
\end{figure}
\begin{figure*}\centering
 \subfigure[] {
    \label{figa:fringes-unit-vis}
    \includegraphics[width=0.28\textwidth]{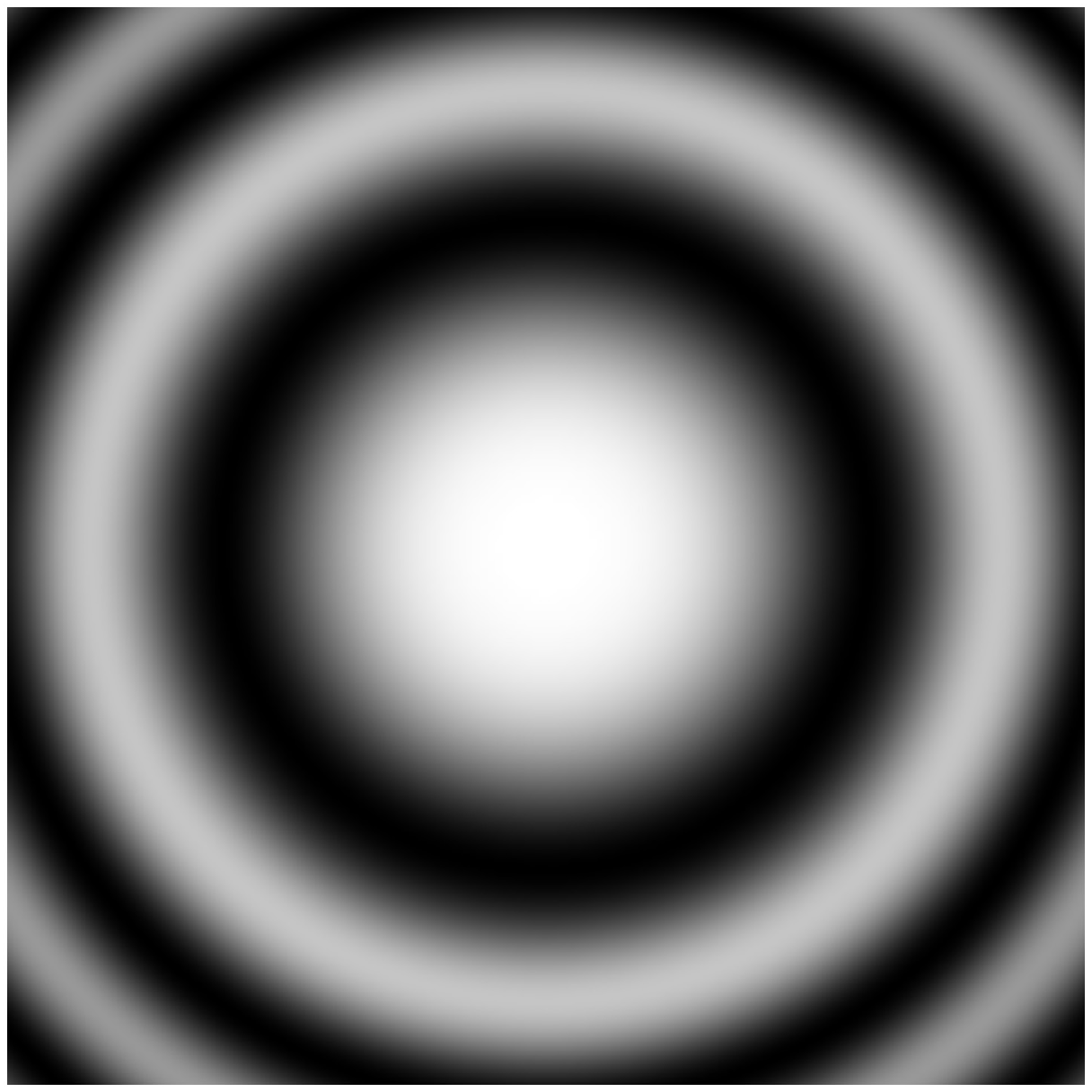}
} 
   \subfigure[] {
    \label{figb:plot-unit-vis}
    \includegraphics[width=0.3\textwidth]{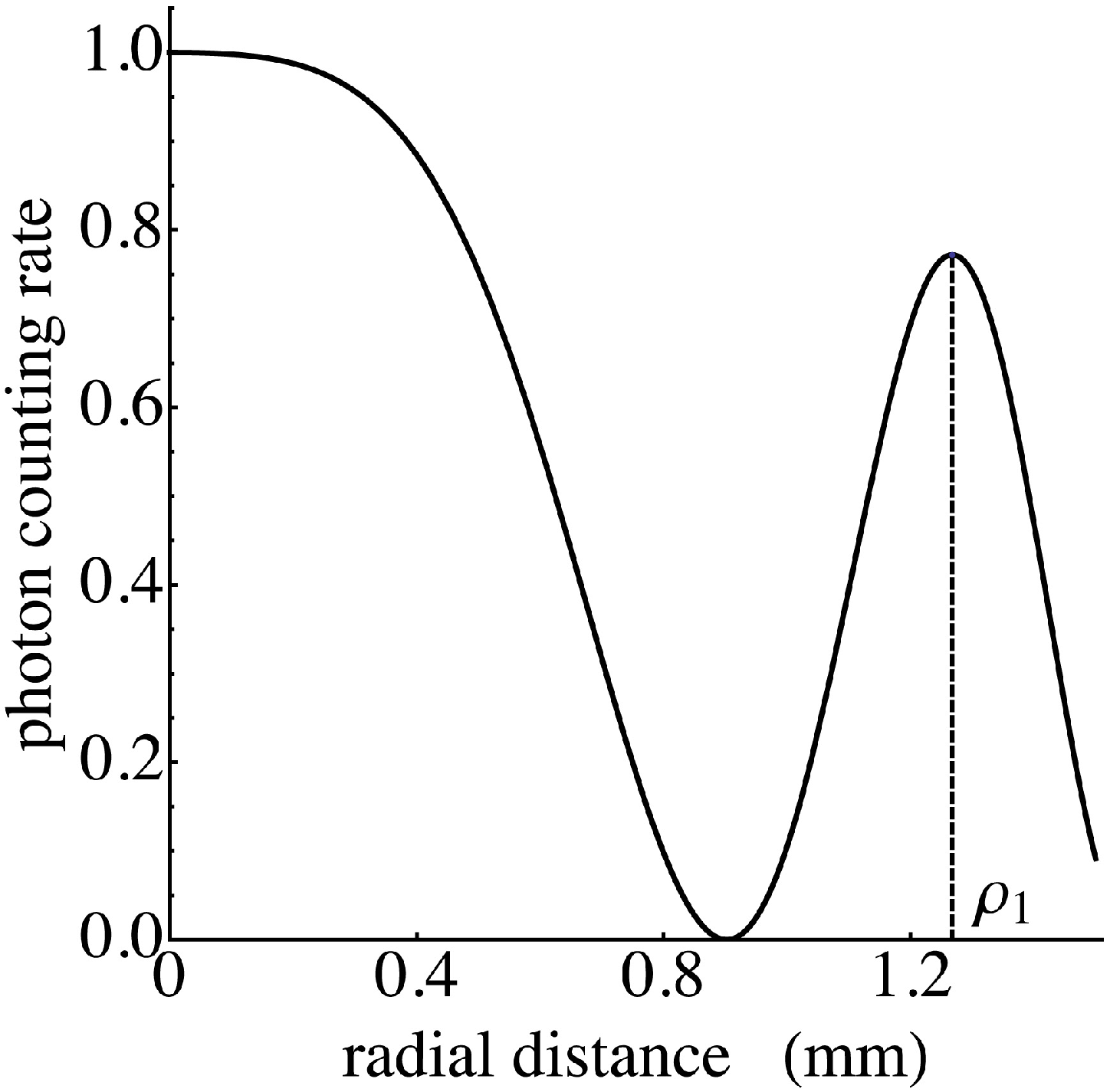}
} 
 \subfigure[] {
    \label{figc:max-d-plot}
    \includegraphics[width=0.3\textwidth]{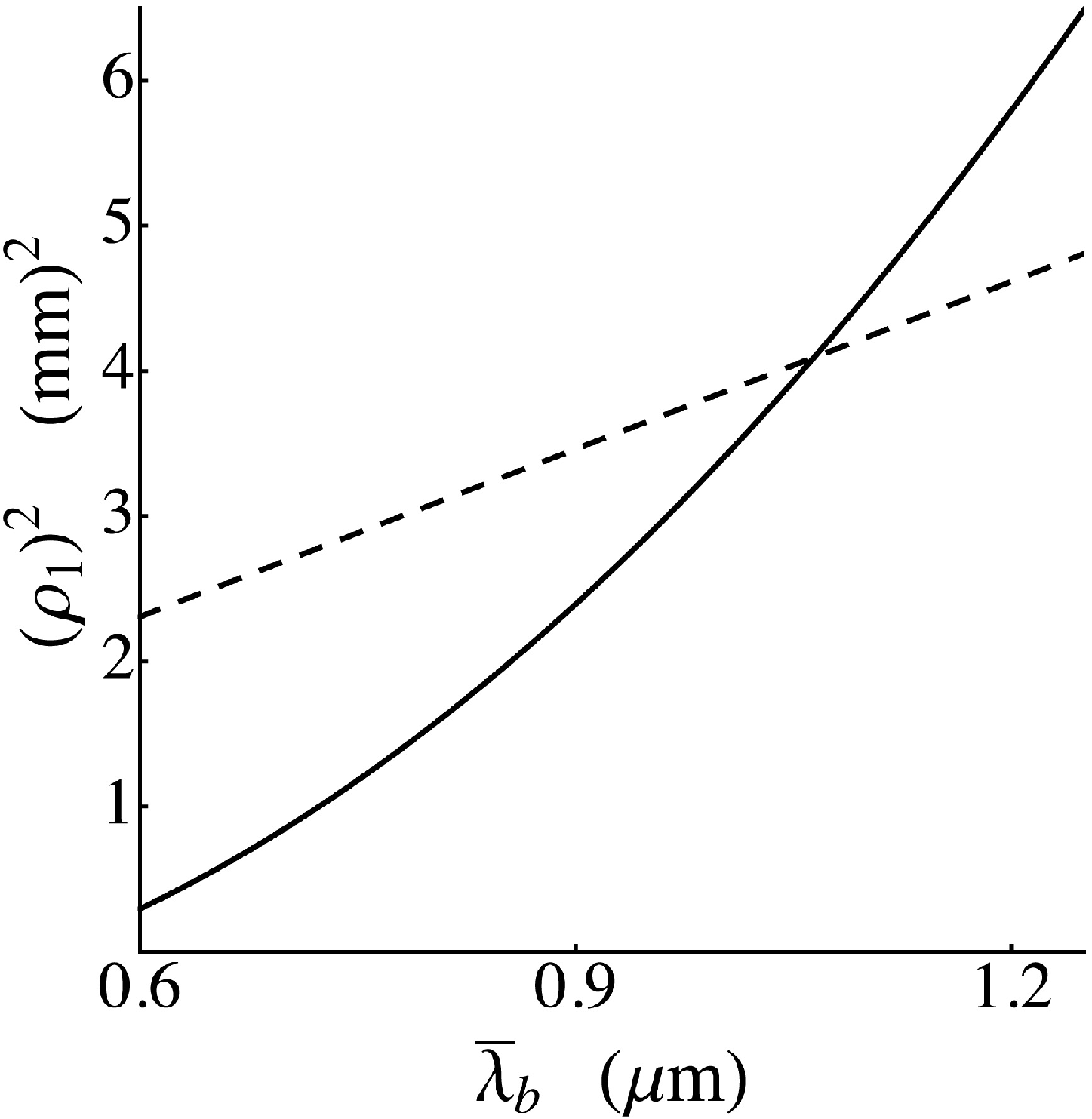}
} 
   \subfigure[] {
    \label{figa:fringes-less-vis}
    \includegraphics[width=0.28\textwidth]{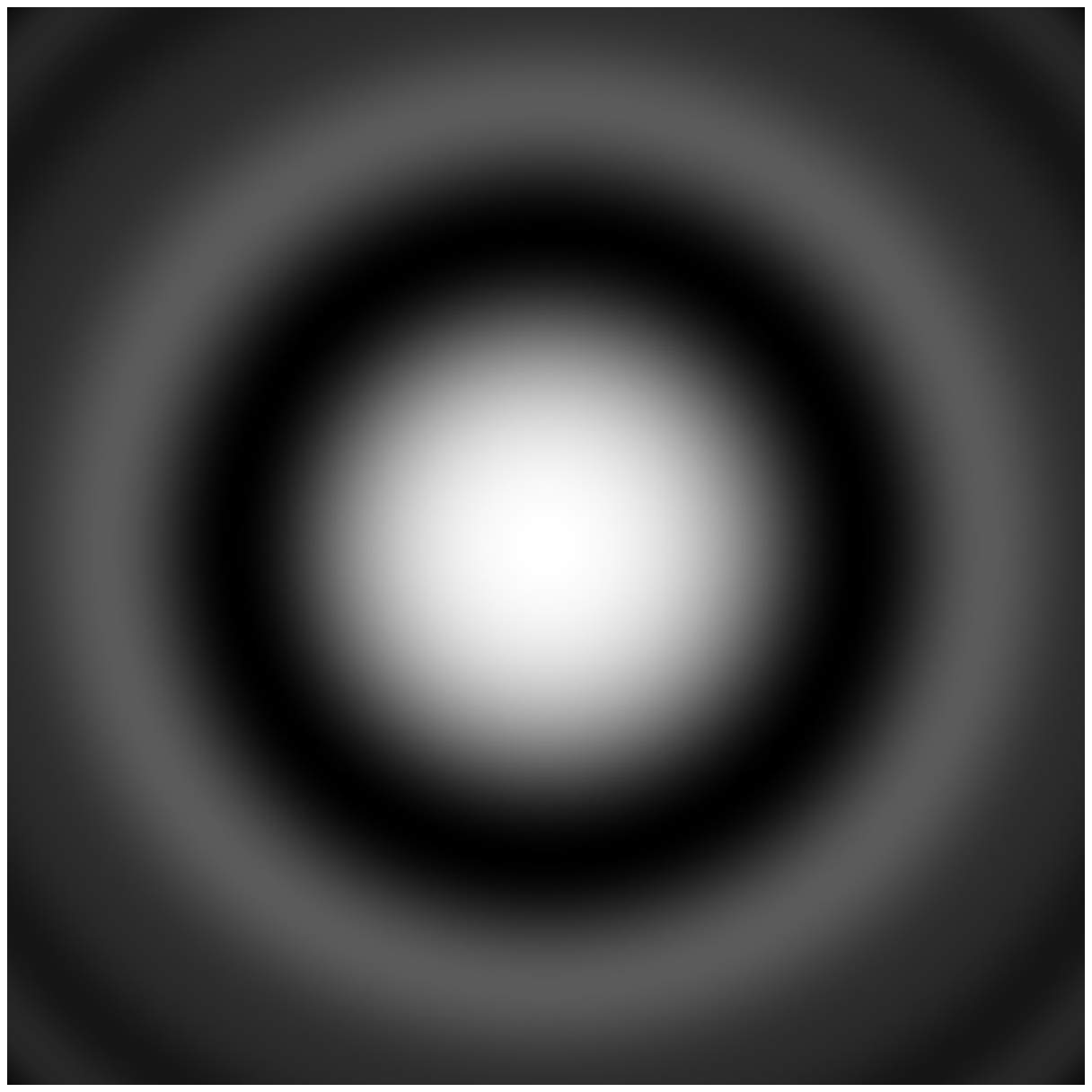}
} 
   \subfigure[] {
    \label{figb:plot-less-vis}
    \includegraphics[width=0.3\textwidth]{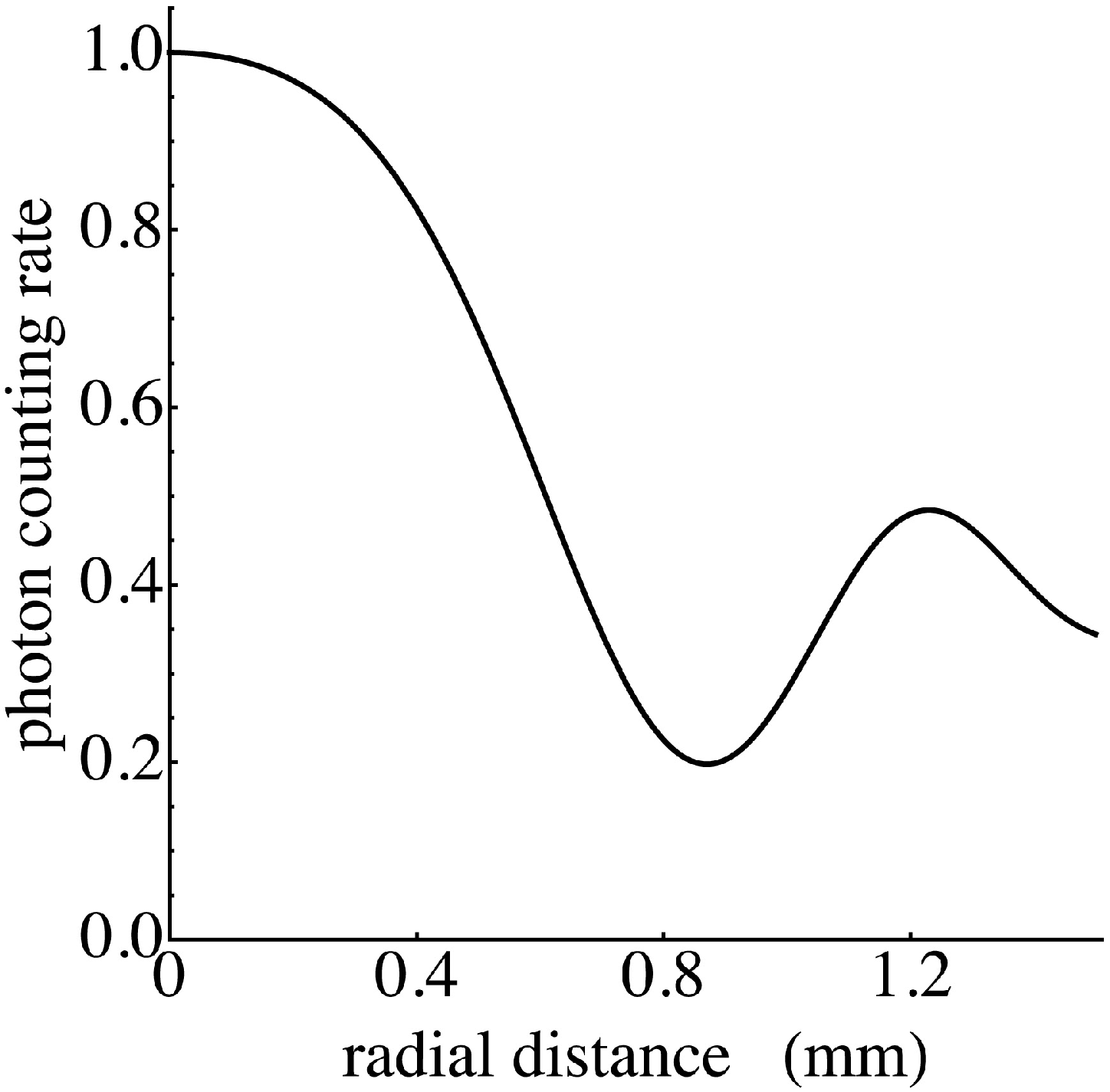}
} 
   \subfigure[] {
    \label{figc:HWHM-less-vis}
    \includegraphics[width=0.3\textwidth]{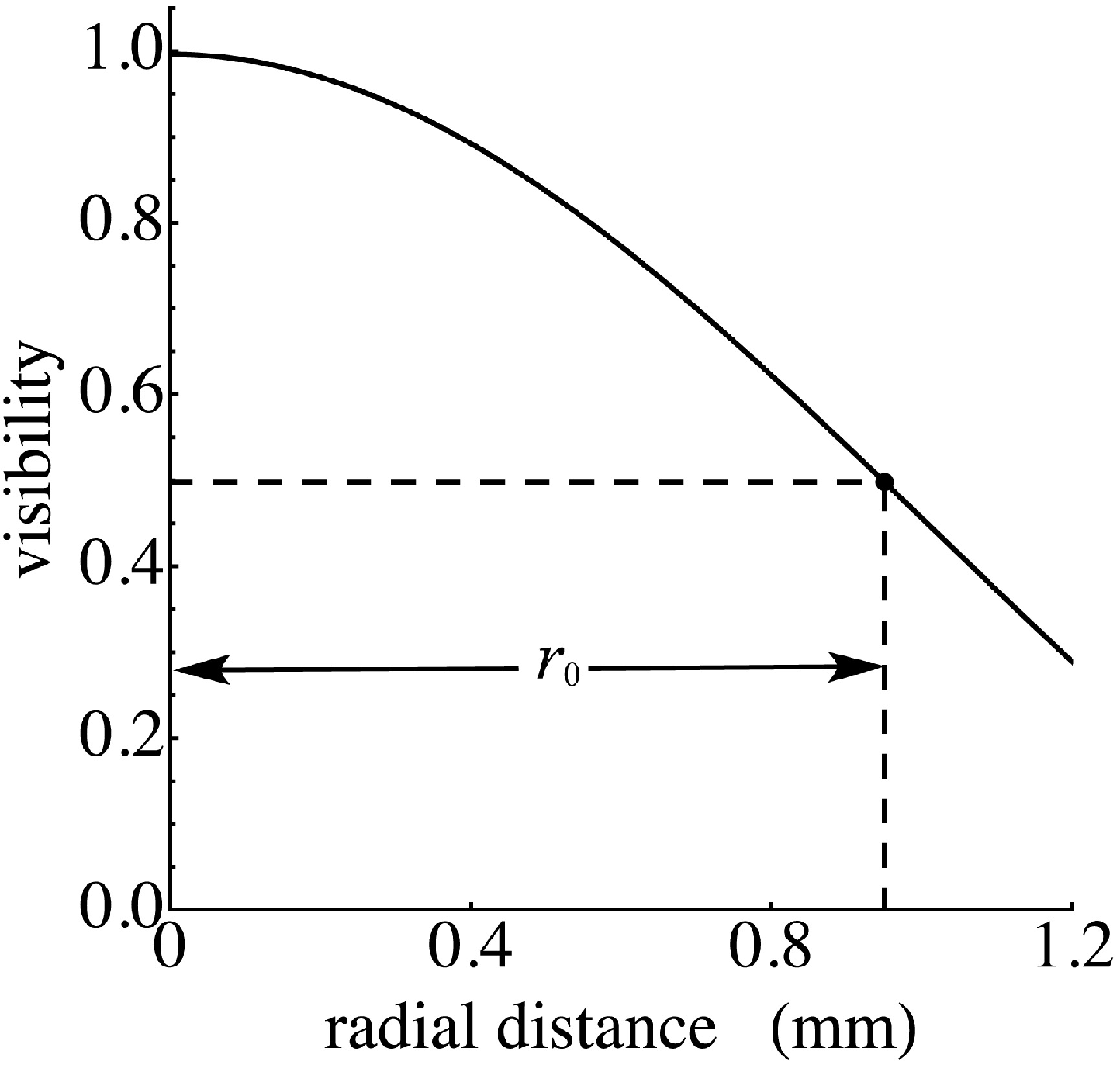}
} \caption{Computationally obtained interference patterns when
momenta of photons $a$ and $b$ are maximally correlated [(a), (b),
(c)] and partially correlated [(d), (e), (f)]. Maximum correlation:
(a) Circular fringes with maximum visibility shown on a 3 mm
$\times$ 3 mm screen for $d_a=1.17$ cm and $\phi_0=0$. (b)
Dependence of the normalized photon counting rate on
$\rho_{k_b}\equiv |\pmb{\rho}_{\k_b}|$ for the fringe pattern. The
minimum value of the intensity is zero implying unit visibility. The
radius of first ($N=1$) bright ring (maximum) is $\rho_1$. (c)
Square of the radius, $(\rho_1)^2$, of the first bright ring plotted
against the wavelength, $\lambda_b$, of the interfering light for
our experiment (solid line) and for a Michelson interferometer
(dashed line). Partial correlation: (d) Fringes of partial
visibility shown on a 3 mm $\times$ 3 mm screen for $d_a=1.17$ cm
and $\phi_0=0$. (e) Dependence of the normalized photon counting
rate on $\rho_{k_b}$ for the fringe pattern. Minimum value of the
intensity is bigger than 0 implying less than unit visibility. (f)
Visibility, $\mathcal{V}(\pmb{\rho}_{\k_b})$, obtained by varying
$\phi_0$ at each point on the fringe pattern [shown in (d)] is
plotted against the distance from the center of the pattern. The
visibility at the center of the pattern: $\mathcal{V}(0)=0.996$. The
HWHM, $r_0$, is the radial distance where the visibility drops to
half of its value at $\rho_{k_b}=0$.}
\label{fig:fringes-computed-par}
\end{figure*}

To illustrate the phenomenon we assume that photons $a$ and $b$ are
emitted into collinear or near-collinear beams and
$\mathbf{f}(\k_b)= \k_0-\k_b$, where $\k_0$ is a constant vector
along the common axis of the beams of $a$- and $b$-photons (Fig.
\ref{fig:th-a-b-rel-ed}.
It now readily follows that $|\k_a|\sin\theta_a=|\k_b|\sin\theta_b$,
where $\theta_a$ and $\theta_b$ are the angles made by $\k_a$ and
$\k_b$, respectively, with $\k_0$ \cite{Note-crys-ref}. In the small
angle limit we then have $\bar{\omega}_a\theta_a\approx
\bar{\omega}_b\theta_b$. Since $|\pmb{\rho}_{\k_b}|=f_0\tan
\theta_b\approx f_0\theta_b$, we obtain the relation $
\theta_a\approx
\bar{\omega}_b|\pmb{\rho}_{\k_b}|/(f_0\bar{\omega}_a)$, where $f_0$
is the focal length of L0 (Fig. \ref{fig:det-sys}). Equation
(\ref{R-at-cam-a-fringe-sp-max-ent-red}) now reduces to the form
\begin{align}\label{R-at-cam-a-fringe-sp-max-ent-red-2}
\mathcal{R}(\pmb{\rho}_{\k_b})\propto P(\k_b) \{1
+\cos[\frac{\bar{\omega}_b^2}{\bar{\omega}_a}
\frac{n(\bar{\omega}_a) d_a}{2cf_0^2}|\pmb{\rho}_{\k_b}|^2]\}.
\end{align}
If $P(\k_b)$ only depends on $\theta_b$, the circular symmetry of
Eq. (\ref{R-at-cam-a-fringe-sp-max-ent-red-2}) suggests that the
fringes are circular in shape and the minimum value of the photon
counting rate across the fringe pattern is zero. Figures
\ref{figa:fringes-unit-vis} and \ref{figb:plot-unit-vis} show the
computationally obtained fringe pattern for the following choices of
expressions and parameters:
$P(\k_b)=\exp[-2\theta_b^2/\sigma_b^2]=\exp[-2\rho_{k_b}^2/(f_0\sigma_b)^2]$,
where $\sigma_b=2.36\times 10^{-2}$ and $f_0=15$ cm; we choose
$n(\bar{\omega}_a)=1$, $\bar{\lambda}_a=1550$ nm and
$\bar{\lambda}_b=810$ nm.
\par
Equation (\ref{R-at-cam-a-fringe-sp-max-ent-red-2}) implies that the
condition of a maximum is given by
\begin{align}\label{max-cond-a-fringes}
\frac{\bar{\omega}_b^2}{\bar{\omega}_a} \frac{n(\bar{\omega}_a)
d_a}{2cf_0^2}(\rho_N)^2=2N\pi,\quad
N=0,1,2,\dots,
\end{align}
where $\rho_N$ is the radius of the $N$-th bright ring and $N=0$
characterizes the central maximum. If the medium between $Q_1$ and
$Q_2$ is nondispersive, it immediately follows that
\begin{align}\label{max-pos-a-fringes}
\rho_N\propto \sqrt{N}~(\bar{\lambda}_b^2/\bar{\lambda}_a)^{1/2}=
\sqrt{N}~\bar{\lambda}_{eq}^{1/2},
\end{align}
where $\bar{\lambda}_a$ and $\bar{\lambda}_b$ are mean wavelengths
of $a$ and $b$ photons, respectively, and
$\bar{\lambda}_{eq}=\bar{\lambda}_b^2/\bar{\lambda}_a$. Clearly, the
radius of the $N$th bright ring is proportional to $\sqrt{N}$ just
like the fringes of equal inclination produced in a Michelson
interferometer (\cite{BW}, Sec. 7.5.4). However, in contrast to the
Michelson interferometer the fringe shift associated with a change
in $d_a$ is \emph{not} characterized by the wavelength of the
interfering light but by an equivalent wavelength
$\bar{\lambda}_{eq}=\bar{\lambda}_b^2/\bar{\lambda}_a$. We
illustrate this fact in Fig. \ref{figc:max-d-plot} by plotting the
square of the radius [$(\rho_1)^2$] of the first ($N=1$) bright ring
(maximum) against the wavelength of the interfering light
($\lambda_b$) and comparing it with the case of a traditional
Michelson interferometer.
\par
It is possible to determine the wavelength $\bar{\lambda}_{eq}$ from
the fringe shift associated with change in $d_a$. The value of
$\bar{\lambda}_a$ can then be obtained if the value of
$\bar{\lambda}_b$ is known. Note that photon $a$ is not detected;
the fringe pattern is obtained by detecting photon $b$ only.
\emph{This implies that one can determine the mean wavelength of
photon} $a$ \emph{without detecting it}.
\par
Equation (\ref{R-at-cam-a-fringe-sp-max-ent}) shows that \emph{the
visibility of the fringes is equal to unity for perfect momentum
correlation}. Although a perfect correlation can be achieved only in
an idealized situation, photon pairs highly correlated in momenta
are now regularly generated in laboratories.

\subsection{No Momentum Correlation}\label{subsec:no-corr}
\par
We now consider the case in which the momenta of the twin-photons
generated by each source are not correlated. As already mentioned in
Sec. \ref{sec:back-theory}, one now has
\begin{align}\label{cond-no-corr}
|C_{\k_{a},\k_{b}}|^2 =P(\k_a)P(\k_b).
\end{align}
Equation (\ref{R-at-cam-a-fringe-sp}) now reduces to
\begin{align}\label{R-at-cam-a-fringe-no-ent}
\mathcal{R}(\pmb{\rho}_{\k_b})&\propto P(\k_b)\sum_{\k_a}P(\k_a)
\big(1+
\cos[\phi_a(\k_a)-\phi_0] \big) \nonumber \\
&=P(\k_b) \times \text{constant}.
\end{align}
In this case, contributions from all $\k_a$ modes get fully averaged
out. It is, therefore, clear that a modulation of $\phi_a(\k_a)$
does not result in the creation of interference fringes. We thus
conclude that when the momenta of photons $a$ and $b$ are
uncorrelated, the visibility of fringes is zero.

\subsection{Partial Momentum Correlation}\label{sebsec:par-corr}
\par
If the transverse momenta of the photons of a pair are partially
correlated, $|C_{\k_{a},\k_{b}}|^2$ can neither be expressed as in
Eq. (\ref{cond-del-corr}) nor as in Eq. (\ref{cond-no-corr}). As a
consequence, the photon counting rate in the camera [Eq.
(\ref{R-at-cam-a-fringe-sp})] can no longer be reduced to a simple
form. However, we can draw some general conclusions. In this case,
the minimum value of intensity at any point of the fringe pattern
can never be zero. The visibility of the fringes must, therefore, be
less than unity. Furthermore, the number of terms contributing to
the sum in Eq. (\ref{R-at-cam-a-fringe-sp}) increases with the range
over which $\k_a$ varies for a given $\k_b$. It thus also follows
that the larger this range, the lower the visibility of the fringes.
\par
To illustrate the phenomenon let us assume that
$\mathcal{P}(\k_{a}|\k_{b})$ is a function of $\k_{a}+\k_{b}\equiv
\k'$. We write $\mathcal{P}(\k_{a}|\k_{b})\equiv |\mu(\k')|^2$ such
that
\begin{align}\label{cond-par-corr}
|C_{\k_{a},\k_{b}}|^2 =P(\k_b)|\mu(\k')|^2.
\end{align}
It now follows from Eq. (\ref{R-at-cam-a-fringe-sp}) that
\begin{align}\label{R-at-cam-a-fringe-par-vis}
&\mathcal{R}(\pmb{\rho}_{\k_b}) \nonumber \\ & \propto P(\k_b)
\sum_{\k'}|\mu(\k')|^2 \left\{1+
\cos\big[\phi_a(\k'-\k_b)-\phi_0\big] \right\}.
\end{align}
We choose $P(\k_b)$ to be given by the same expression as above [see
the text below Eq. (\ref{R-at-cam-a-fringe-sp-max-ent-red-2})]. The
photon counting rate, $\mathcal{R}(\pmb{\rho}_{\k_b})$, is
determined by replacing the summation in Eq.
(\ref{R-at-cam-a-fringe-par-vis}) by an integration (see Appendix 2)
and assuming
\begin{align}\label{cond-prob-form}
\mathcal{P}(\k_{a}|\k_{b})\equiv |\mu(\k')|^2=
\delta(k'-k_0')\exp\left[-2\theta'^2/\sigma_\theta^2\right],
\end{align}
where $k'=|\k'|$, $k_0'$ is a positive constant \cite{Note-k0},
$\delta$ represents the Dirac delta function, $\theta'$ is the angle
made by $\k'$ with the optical axis, and the positive quantity
$\sigma_\theta$ shows how strongly the momenta are correlated
\cite{Note-prob-form}: the higher the value of $\sigma_\theta$, the
weaker the correlation between momenta. Figures
\ref{figa:fringes-less-vis} and \ref{figb:plot-less-vis} illustrate
the computationally obtained fringe pattern for
$\sigma_\theta=9.37\times 10^{-4}$. A comparison between Figs.
\ref{figa:fringes-unit-vis} and \ref{figa:fringes-less-vis} [or
between Figs. \ref{figb:plot-unit-vis} and \ref{figb:plot-less-vis}]
shows that the visibility of the fringes is reduced when the momenta
of photons $a$ and $b$ are less correlated.
\begin{figure*}\centering
 \subfigure[] {
    \label{figa:vis-center-sig-plot}
    \includegraphics[width=0.4\textwidth]{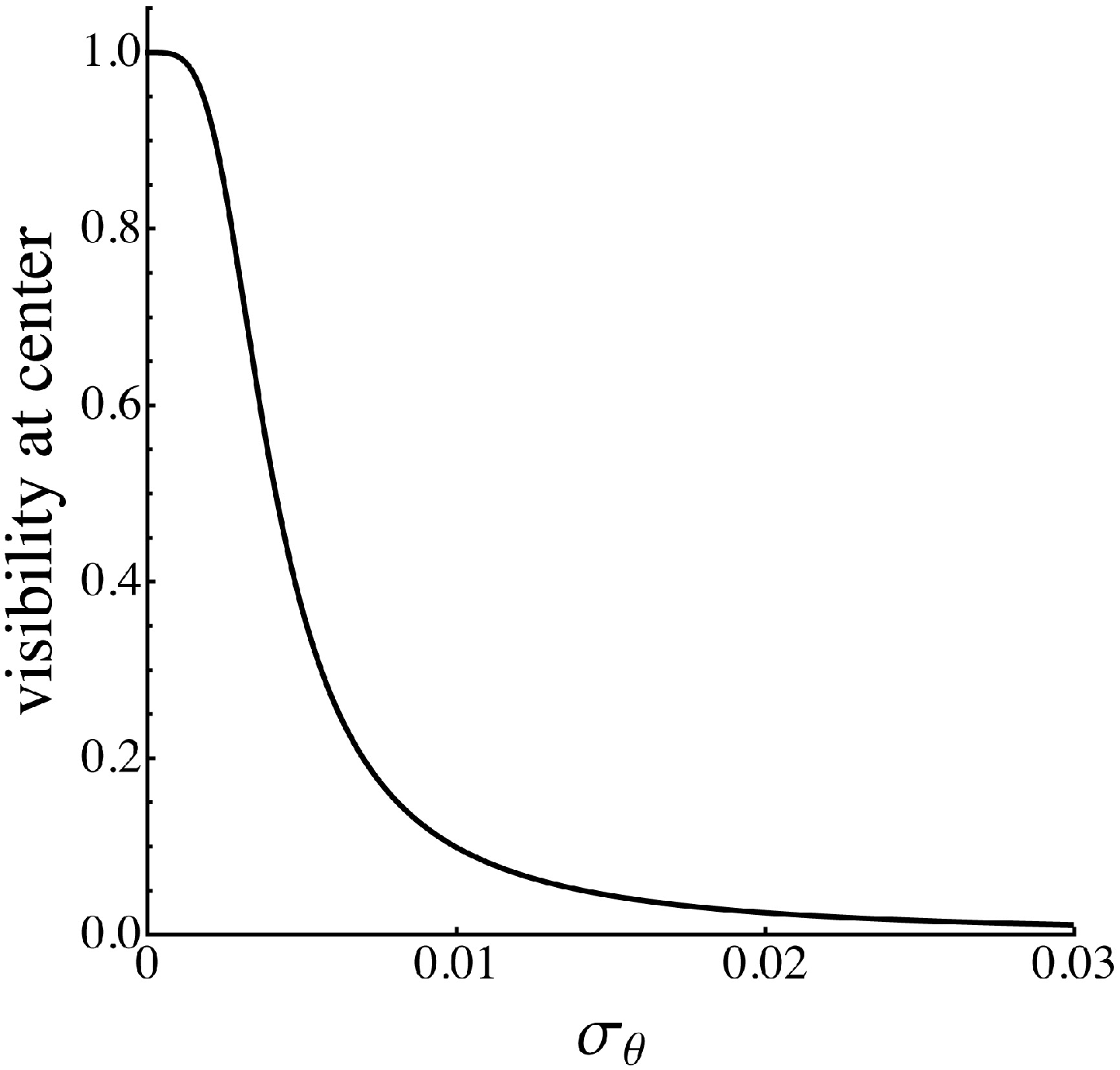}
} 
   \subfigure[] {
    \label{figb:vis-HWHM-drop}
    \includegraphics[width=0.38\textwidth]{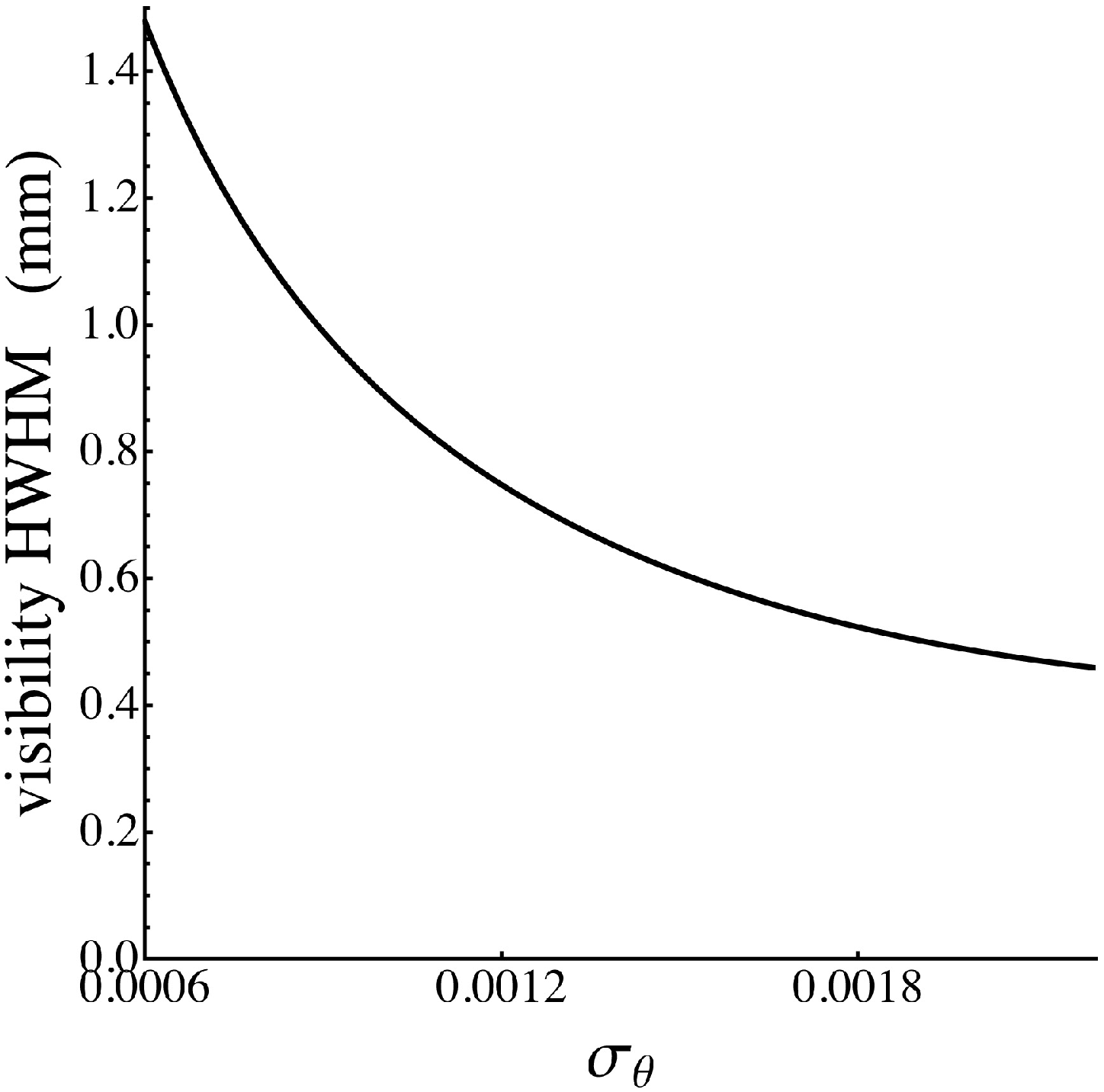}
} \caption{Drop of visibility with decreasing momentum correlation,
i.e., with increasing value of $\sigma_\theta$. (a) Visibility at
the center of the pattern [$\mathcal{V}(0)$] drops as
$\sigma_\theta$ increases. (b) The HWHM, $r_0$, [shown in Fig.
\ref{figc:HWHM-less-vis}], is plotted against $\sigma_\theta$.
Higher value of $\sigma_\theta$ implies lesser momentum correlation
and higher value of $r_0$ implies higher visibility. Chosen
parameters: $d_a=1.17$ cm, $\bar{\lambda}_a=1550$ nm,
$\bar{\lambda}_b=810$ nm, and $\sigma_b=1.67\times 10^{-2}$.}
\label{fig:vis-drop}
\end{figure*}
\par
We further investigate the relationship between the visibility of
fringes and the momentum correlation by determining the visibility
at each point on the fringe pattern. Note that $\phi_0$ is
independent of $\k_a$ and can be changed while $\phi_b(\k_b)$ is
fixed. The maximum ($\mathcal{R}_{\text{max}}$) and minimum
($\mathcal{R}_{\text{min}}$) values of the photon counting rate at
each point are determined by varying $\phi_0$ and the visibility is
obtained by formula (\ref{vis-def}). We find that the visibility at
each point is given by (see Appendix 2)
\begin{align}\label{vis-at-a-pt}
\mathcal{V}(\pmb{\rho}_{\k_b})&=\frac{1}{\gamma}
\exp\left(-\sigma_\theta^2|\pmb{\rho}_{\k_b}|^2/\chi^2 \right)
\nonumber \\& ~~\times
\left|D_{-2}\left\{|\pmb{\rho}_{\k_b}|g(\sigma_\theta)\right\}
+D_{-2}\left\{-|\pmb{\rho}_{\k_b}|g(\sigma_\theta)\right\}\right|,
\end{align}
where $A=\pi d_a \bar{\lambda}_a/(f_0\bar{\lambda}_b)^2$,
$B=f_0\bar{\lambda}_b k_0'/(2\pi)$,
$\gamma=\sqrt{4+\sigma_\theta^4A^2B^4}$, $\chi=\gamma/(AB)$,
$g(\sigma_\theta)
=i\sqrt{2}AB\sigma_\theta/\sqrt{2-i\sigma_\theta^2AB^2}$, and $D_n$
is the parabolic cylinder function of order $n$. Figure
\ref{figc:HWHM-less-vis} illustrates the dependence of the
visibility on the distance from the center of the pattern. The drop
of visibility with the radial distance is characterized by the half
width at half maximum (HWHM), i.e., by the distance, $r_0$, from the
center of the pattern, at which the visibility drops to the half of
its value at the center.
\par
Note that the maximum value of the visibility is obtained at
$\pmb{\rho}_{\k_b}=0$, i.e., at the center of the fringe pattern. It
follows from Eq. (\ref{vis-at-a-pt}) that
\begin{align}\label{vis-at-center}
\mathcal{V}(0)=\frac{2}{\gamma}=\frac{2}{\sqrt{4+\sigma_\theta^4A^2B^4}},
\end{align}
where we have used the fact $D_{-2}(0)=1$. Clearly, when
$\sigma_\theta \rightarrow 0$, i.e., when the momentum correlation
is maximum, $\mathcal{V}(0)=1$. In the other extreme case, when
$\sigma_\theta \rightarrow \infty$, i.e., when the momentum
correlation attains the minimum value, $\mathcal{V}(0)=0$. When
$\sigma_\theta$ has a finite non-zero value, i.e., when the momenta
of photons $a$ and $b$ are partially correlated, one has
$0<\mathcal{V}(0)<1$. Figure \ref{figa:vis-center-sig-plot} shows
the dependence of $\mathcal{V}(0)$ on $\sigma_\theta$. We further
show, in Fig. \ref{figb:vis-HWHM-drop}, that the value of  the
visibility HWHM ($\r_0$) decreases as $\sigma_\theta$ increases.
Figures \ref{figa:vis-center-sig-plot} and \ref{figb:vis-HWHM-drop}
thus clearly illustrate that the visibility of the pattern reduces
as the transverse momenta of photons $a$ and $b$ becomes less
correlated.

\section{Determining the Momentum Correlation from the Fringe Pattern}
\label{sec:mom-cor-from-vis} It is clear from the previous section
that the correlation between the transverse momenta of photons $a$
and $b$ governs the visibility of the fringe pattern that is
obtained by detecting photon $b$ only. We now justify that under
reasonable assumptions a quantitative measure of the momentum
correlation between the two photons can be obtained from this
visibility.
\par
Let us first examine the example considered in Sec.
\ref{sebsec:par-corr}. Using the expressions for $A$ and $B$ [see
the text below Eq. (\ref{vis-at-a-pt})], one finds from Eq.
(\ref{vis-at-center}) that
\begin{align}\label{sigma-eval}
\sigma_\theta^2= \frac{8\pi}{k_0'^2\bar{\lambda}_a
d_a}\left(\frac{1}{[\mathcal{V}(0)]^2}-1\right)^{\frac{1}{2}}.
\end{align}
It is thus clear that $\sigma_\theta$ can be uniquely determined
from the visibility of the interference pattern. Since the
conditional probability [see Eq. (\ref{cond-prob-form})] is given by
$\mathcal{P}(\k_{a}|\k_{b}) \propto
\exp\left[-2\theta'^2/\sigma_\theta^2\right]$, it can be immediately
determined once $\sigma_\theta$ is known. (The conditional
probability is often measured in the procedures that involve
heralded detection \cite{HBBB-sp-ent-bell-viol,Padgett-sp-ent-cam}.)
The quantity $P(\k_b)$ is directly obtained from the spatial
dependence of the normalized photon counting rate in the camera when
only one of the beams ($b_1$ or $b_2$) of photon $b$ is detected.
Now using Eq. (\ref{cond-prob}), one can determine the joint
probability $P(\k_a,\k_b)$ that governs the momentum correlation
between the two photons $a$ and $b$. Alternatively, the conditional
probability can also be determined from the fact that the value of
the visibility HWHM ($r_0$) reduces with decreasing momentum
correlation (Fig. \ref{figb:vis-HWHM-drop}).
\par
We stress that the method of determining the joint probability
(density) is not restricted to this particular example. If the two
photons are not emitted into collinear or near-collinear beams, the
momentum correlation can still be determined from the visibility of
the fringes. In this case, however, the expression for visibility is
no longer given by Eq. (\ref{vis-at-a-pt}), and a more involved
computational technique might be required to determine the
conditional probability. The other assumptions made in our example
are: 1) the photons are propagating in the form of paraxial beams;
2) $\mathcal{P}(\k_{a}|\k_{b})$ is a function of $\k_{a}+\k_{b}$,
i.e., of $\pmb{\mathpzc{p}}_a+\pmb{\mathpzc{p}}_b$; and 3)
$\mathcal{P}(\k_{a}|\k_{b})$ has a Gaussian form. Most traditional
methods of determining momentum correlation usually require these
three assumptions to be made. Note that our method applies to more
general situations. When assumption 1 does not hold, the correlation
between the longitudinal components of momenta also contributes to
the visibility. However, a more involved detection system can in
principle be employed to determine the correlation between the
three-dimensional momenta of the two photons. Assumption 2 holds in
many practical situations (see, for example,
\cite{TGTSW-prop-int-corr}). This assumption or an equivalent one
might be necessary for determining $\mathcal{P}(\k_{a}|\k_{b})$.
This is because Eq. (\ref{R-at-cam-a-fringe-sp}) reduces to a three
dimensional Fredholm integral equation of the first kind (in the
continuous variable limit) only under such an assumption. This
integral equation is uniquely solvable in principle; furthermore, in
many cases its dimensionality can be reduced due to symmetries
present in the system. It thus follows that a specific functional
form of $\mathcal{P}(\k_{a}|\k_{b})$ does not need to be assumed,
i.e., assumption 3 is not essential.

\section{Comparison with Experimental
Results}\label{sec:exp-confirm}

We have experimentally verified the above mentioned results
\cite{HLLLZ-eq-wvln,HLLLZ-mom-corr-exp}. Here, we make a brief
instructive comparison of our theoretical predictions with the
experimental observations.
\par
In the experiments, nonlinear crystals (ppKTP) pumped by mutually
coherent laser beams have been used as biphoton sources. Each
crystal can produce a photon pair ($a$, $b$) by the process of
spontaneous parametric down-conversion. In this case, the wave
vector, $\k'=\k_a+\k_b$, represents a wave vector of the pump.
\par
The momentum correlation between the two photons is modulated by the
tightness of focusing of the pump beam into the crystals. The
tighter the focus of the pump, the bigger is the range over which
$\k'$ can vary for a particular choice of $\k_b$, i.e., the value of
$\sigma_{\theta}$ increases (see Eq. (\ref{cond-prob-form})). For a
Gaussian pump waist $w_p$, one has
$\sigma_{\theta}=\bar{\lambda}_p/(\pi w_p)$, where
$\bar{\lambda}_p=2\pi/k_0'$ is the mean wavelength of the pump beam
(see endnote \cite{Note-k0}). The experimental values of
$\bar{\lambda}_p$, $\bar{\lambda}_a$, and $\bar{\lambda}_b$ are
$532$ nm, $1550$ nm, and $810$ nm, respectively.
\par
To achieve a high quality alignment of the beams of photon $a$, a
$4f$ lens system was placed between the two sources (crystals) on
the path of the $a$-photon beam. When the $4f$ lens system is fully
balanced, it images the first source ($Q_1$) on the second source
($Q_2$). In this case, the effective propagation distance, $d_a$,
between the two sources becomes zero. Nonzero values of $d_a$ were
obtained by unbalancing the $4f$ lens system (for further details
see \cite{HLLLZ-eq-wvln}).
\par
\begin{figure}[t]  \centering
\includegraphics[width=0.4\textwidth]{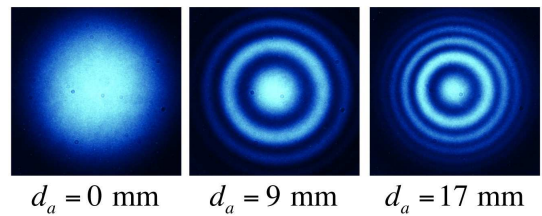}
\qquad \caption{(Adapted from \cite{HLLLZ-eq-wvln}.) Experimentally
observed fringes for different values of the effective propagation
distance ($d_a$), when the pump beams were highly collimated. The
equivalent wavelength, $\bar{\lambda}_{eq}$, which characterizes the
fringe shift associated with $d_a$ was experimentally determined and
found to be $420\pm 7$ nm; the theoretically predicted value is
$423$ nm. One does not need to know the wavelengths of the pump,
$a$, and $b$ photons for the experimental determination of
$\bar{\lambda}_{eq}$.} \label{figa:exp-fringes-max-vis}
\end{figure}

Figure \ref{figa:exp-fringes-max-vis} shows fringe patterns obtained
for different values of $d_a$, when the pump is highly collimated
(very weakly focused at the crystals). The consequent high momentum
correlation between photons $a$ and $b$ results in fringes with high
visibility as suggested by the theoretical analysis. The equivalent
wavelength, $\bar{\lambda}_{eq}$, has been determined experimentally
and found to be $420\pm 7$ nm, where the theoretically predicted
value is approximately $423$ nm.
\par
\begin{figure}[htbp]  \centering
\includegraphics[width=0.4\textwidth]{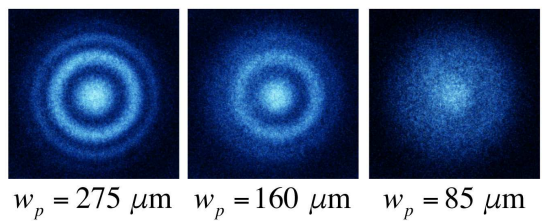}
\qquad \caption{(Adapted from \cite{HLLLZ-mom-corr-exp}.)
Experimentally observed fringes for different values of the pump
waist at the crystals and a fixed value of $d_a=11.7$ mm. The values
of $\sigma_{\theta}$ corresponding to the pump waists (left to
right) are $6.16\times 10^{-4}$, $1.06\times 10^{-3}$, and
$1.99\times 10^{-3}$, respectively. The fringe visibility drops as
the momentum correlation between the two photons reduces.}
\label{figb:exp-fringes-par-vis}
\end{figure}

Figure \ref{figb:exp-fringes-par-vis} shows that the fringes blur
out as the pump beams are more tightly focused at the crystals. It
illustrates that when the momentum correlation between the two
photons reduces, the visibility also reduces. In the experiment, the
dependence of the visibility on the distance from the center of the
pattern was measured for different values of $w_p$.
It was found that the visibility drops with the radial distance as
suggested by Fig. \ref{figc:HWHM-less-vis}. The experimentally
obtained dependence (\cite{HLLLZ-mom-corr-exp}, Fig. 2B) of the
corresponding FWHM ($2r_0$) on $w_p$ matches with the theoretical
dependence predicted by Eq. (\ref{vis-at-a-pt}). The variance of the
conditional transverse momentum correlation was also determined from
the visibility of the fringes; for further details see
\cite{HLLLZ-mom-corr-exp}.

\section{Summary and Conclusion}\label{sec:conc}
We have shown that by using two spatially separated identical
biphoton sources a novel single-photon fringe pattern can be
created. We have restricted the analysis to circular fringes that
resemble fringes of equal inclination (\cite{BW}, Sec. 7.5.1). It is
not difficult to envision that fringes of equal thickness
(\cite{BW}, Sec. 7.5.2) can also be created in our system: this can
be done by a slight misalignment of the beams $a_1$ and $a_2$ when
the distance between the sources is very small. Our system can
therefore produce interference fringes that resemble the fringes
observed in a Michelson interferometer (\cite{BW}, Sec. 7.5.4).
However, in contrast to our system, a Michelson interferometer
superposes two beams that are created from a single beam by division
of amplitude. In our case, the interfering beams are generated
separately by two sources that produce photon pairs. Furthermore,
the interference fringes produced in our system have some novel
properties. These fringes are obtained by detecting only one photon
of each photon pair. We have discussed in Sec. \ref{subsec:max-corr}
that the fringes shift when the optical path traversed by the
undetected photon changes. This fringe shift is characterized by a
combination of wavelengths of both photons, which we experimentally
demonstrate in Ref. \cite{HLLLZ-eq-wvln}. This phenomenon allows us
to determine the wavelength of a photon without detecting it.
\par
A striking feature of the fringe pattern is that the visibility
decreases with decreasing correlation between the transverse momenta
of each photon pair. This observation opens up a new avenue for
measuring correlation between two quantum entities. As shown in Sec.
\ref{sec:mom-cor-from-vis}, this property of the fringes can be used
to determine the momentum correlation. The result is remarkable
because most traditional methods of measuring any sort of
correlation between photons of a pair involve coincidence or an
equivalent detection technique (see, for example,
\cite{HBBB-sp-ent-bell-viol,Padgett-sp-ent-cam,HALRW-sp-ent-PRL-14,PME-tr-md-ent-PRA-09,Silberhorn-g2-corr-exp-OL,HH-high-d-ent-13,JCCL-trans-ent-15,TRGTW-coarse-grain-PRL-13}).
In our method, no coincidence or heralded detection or post
selection is required \cite{Note-Christ-paper}; only measuring the
single-photon counting rate is enough. Since we do not need to
detect one of the photons of the pair, the method allows us to
access wavelengths for which good detectors are not available. This
fact also extends the experimental reach.
\par
Measurement of the correlation between transverse momenta of two
photons is essential for the verification of spatial entanglement.
Furthermore, we believe that our method can be generalized to
measure position correlation between two photons and also to other
degrees of freedom (for example, one can apply it to measure
spectral correlation of biphotons). Our results therefore open up
the possibility of developing a novel method of verifying
entanglement without coincidence or heralded detection
\cite{Note-ent-pure}. Since entanglement plays a vital role in
fundamental tests of quantum mechanics
\cite{EPR,Bohm-hid-var-1,Bohm-hid-var-2,Bell-ineq-orig} and has
important applications in quantum information and communication
science (see, for example, \cite{HHHH-ent-RMP-2009}), we expect that
this direction of research will have a broad significance in the
future.
\par
Finally, recent experimental developments in the fields of microwave
superconducting cavities \cite{LPHH-mandel-MSC}, trapped ions
\cite{Blatt-Wineland-Nat-2008}, atomic systems
\cite{Keller-Z-PRA-2014,Aspect-at-HOM-2015}, and superconducting
circuits \cite{B-etal-Nat-2014} shows the possibility of
generalizing our method to other quantum mechanical entities.

\section*{Acknowledgments}\label{sec:ackn}
We thank Dr. Fabian Steinlechner for useful discussions. This work
was supported by the Austrian Academy of Sciences (\"OAW- IQOQI,
Vienna), and the Austrian Science Fund (FWF) with SFB F40 (FOQUS)
and W1210-2 (CoQus). R.L. was supported by National Science Centre
(Poland) grants 2015/17/D/ST2/03471, 2015/16/S/ST2/00424, the Polish
Ministry of Science and Higher Education, and the Foundation for
Polish Science (FNP).


\section*{Appendix 1}\label{sec:append-1}
We show the derivation of Eq. (\ref{state-sup-prod-mm}).
\par
It follows from Eq. (\ref{state-two-ph-mm}) that the by photon state
generated by each source can be represented as
\begin{align}\label{state-each-source}
&\ket{\psi_j}=\sum_{\k_{a_j},\k_{b_j}}C_{\k_{a},\k_{b}}\ket{\k_{a_j}}_{a_j}
\ket{\k_{b_j}}_{b_j} \nonumber
\\&=\sum_{\k_{a},\k_{b}}\left[C_{\k_{a_j},\k_{b_j }}
\opa_{a_j}^{\dag}(\k_{a_j})
\opa_{b_j}^{\dag}(\k_{b_j})+\text{h.c.}\right]\ket{0}_{a_j}
\ket{0}_{b_j},
\end{align}
where $j=1,2$ label the sources, $\ket{0}$ represents a vacuum
state, and  $\opa_{a}^{\dag}$ and  $\opa_{b}^{\dag}$ represents
creation operators of photons $a$ and $b$, respectively, and h.c.
represents Hermitian conjugate. When beam $a_1$ is sent through
source $Q_2$ and is aligned with beam $a_2$, one has for each mode
$\k_a$
\begin{align}\label{alignment-cond}
\opa_{a_2}(\k_{a})=\exp[i \phi_a(\k_a)]\opa_{a_1}(\k_{a}),
\end{align}
where $\phi_a(\k_a)$ is the phase acquired by the plane-wave mode
$\k_a$ due to propagation from $Q_1$ to $Q_2$.
\par
Since photon pairs emitted by both sources are never simultaneously
present in out system, the quantum state of light is obtained by a
linear superposition of the states generated by each source with the
condition imposed by Eq. (\ref{alignment-cond}). Equation
(\ref{state-sup-prod-mm}) thus immediately follows from Eqs.
(\ref{state-each-source}) and (\ref{alignment-cond}).

\section*{Appendix 2}\label{sec:append-2}
We discuss some mathematical steps used in Sec.
\ref{sebsec:par-corr} following Eq.
(\ref{R-at-cam-a-fringe-par-vis}).
\par
Recall that $\theta'$, $\theta_a$, and $\theta_b$ are the angles
made by $\k'$, $\k_a$, and $\k_b$, respectively, with the optical
axis; $\theta'$, $\theta_a$, and $\theta_b$ are very small angles.
Using the fact that the change in $\phi_a$ is due to change in
propagation distance $d_a$, $\phi_a$ is expressed in terms of $d_a$
and $\theta_a$ as shown in Eq. (\ref{ph-ch-a-fringes}) of Sec.
\ref{subsec:max-corr}; $\theta_a$ is further expressed in terms of
$\theta'$ and $\theta_b$ using the condition $\k'=\k_{a}+\k_{b}$.
Substituting for $\phi_a$ into Eq.
(\ref{R-at-cam-a-fringe-par-vis}), using the assumed forms of
$P(\k_b)$ and $|\mu(\k')|$, and replacing the summation by an
integration, we obtain
\begin{align}\label{R-at-cam-a-fringe-par-vis-cont}
\mathcal{R}(\pmb{\rho}_{\k_b})\propto \int_0^{\Delta\theta} &
d\theta' \exp(-2\theta'^2/\sigma_\theta^2)
\exp[-2|\pmb{\rho}_{\k_b}|^2/(f_0\sigma_b)^2]\nonumber
\\ &\times\theta' \big\{
2+\cos[A(B\theta'-|\pmb{\rho}_{\k_b}|)^2-\phi_0]\nonumber \\ &~~~+
\cos[A(B\theta'+|\pmb{\rho}_{\k_b}|)^2-\phi_0]\big\},
\end{align}
where $A=\pi d_a \bar{\lambda}_a/(f_0\bar{\lambda}_b)^2$,
$B=f_0\bar{\lambda}_b k_0'/(2\pi)$, $\Delta\theta$ is the maximum
range up to which $\theta'$ can vary ($\sigma_\theta \ll
\Delta\theta$), and we have used the relations
$P(\k_b)=\exp[-2\theta_b^2/\sigma_b^2]$ and
$|\pmb{\rho}_{\k_b}|^2\approx f_0^2\theta_b^2$. Since $\sigma_\theta
\ll \Delta\theta$, the upper limit of the integration in Eq.
(\ref{R-at-cam-a-fringe-par-vis-cont}) can be replaced by $\infty$,
and the integral can be expressed in terms of standard integrals
whose values are known \cite{GR-ser-int}. An explicit form of
$\mathcal{R}(\pmb{\rho}_{\k_b})$ can thus be obtained and is found
to have the form
\begin{align}
\mathcal{R}(\pmb{\rho}_{\k_b})\propto
1+\mathcal{V}(\pmb{\rho}_{\k_b})\cos\left[\phi_0+\beta(\pmb{\rho}_{\k_b})\right],
\end{align}
where $\mathcal{V}(\pmb{\rho}_{\k_b})$ is given by Eq.
(\ref{vis-at-a-pt}); an explicit form of $\beta(\pmb{\rho}_{\k_b})$
is not required for determining the visibility.



\begin{thebibliography}{99}
\bibitem {Note-cont-var} Since $\k_a$ and $\k_b$ are continuous
variables, the summation is to be interpreted as an integration. We
use the summation symbol for simplicity.
\bibitem {Note-lens-sys} In case the propagation length cannot
be changed appreciably due to experimental constraints, a similar
phase change can be introduced by using an appropriate lens system
or phase shifters.
\bibitem {Note-crys-ref} This situation occurs when the photon
pair is produced by collinear or near-collinear down-conversion
process. In such a case one needs to consider the effect of
refraction at the crystal surface. This, however, does not change
any of our results.
\bibitem {Note-k0} When the photon pair is produced by spontaneous parametric
down-conversion, $k_0'=2\pi/\bar{\lambda}_p$, where
$\bar{\lambda}_p$ is the mean wavelength of the pump photon.
\bibitem {Note-prob-form} The delta function has been used
to approximately represent the very narrow frequency bandwidth of
the pump photon. The Gaussian dependence on $\theta'$ has been
observed expreimentally in several cases (see, for example,
\cite{HBBB-sp-ent-bell-viol,Padgett-sp-ent-cam}).
\bibitem {Note-Christ-paper} Theoretical calculations on probing
multimode squeezing with correlation functions
\cite{Silberhorn-g2-corr-th-NJP} show that it would also be possible
to determine correlation between two photons from the intensity
correlation measured by detecting only one of them. A followup
experiment (\cite{Silberhorn-g2-corr-exp-OL}, Fig. 2(b)) verified
the conclusion, albeit the other photon was also detected and
eventually the ``heralded'' intensity correlation was measured.
\bibitem {Note-ent-pure} Under the assumption that the photon pair
is in a pure state, measurement of the lowest-order spatial
coherence in one beam has been used to verify spatial entanglement
of the pair (see, for example, \cite{JCCL-trans-ent-15}).
\end{thebibliography}
\end{document}